\documentclass[aps,prb]{revtex4}
\usepackage{amsmath}
\usepackage{color}
\usepackage{amssymb}
\usepackage{graphicx}
\usepackage[utf8]{inputenc}
\usepackage{textcomp}
\usepackage{color}
\usepackage{esint}

\begin{document}

\title{ 
Current induced magnetisation in metal without space-inversion symmetry}

\author{V.P.Mineev}
\affiliation{Landau Institute for Theoretical Physics, 142432 Chernogolovka, Russia}

\begin{abstract}
Magneto-electric effect, that is an appearance of magnetisation induced by electric current is allowed by symmetry in metals with crystal structure without space inversion. The microscopic origin of this effect is spin-orbit coupling of electrons with a non-centrosymmetric crystal lattice lifting spin degeneracy of electron energy
and mixing  spin and  orbital degrees of freedom.
The presented  calculation of magnetisation induced by current based on the application of kinetic equation for the matrix distribution function of electrons occupying the states in two bands split by the spin-orbit interaction.

\end{abstract}
\date{\today}
\maketitle

\section{Introduction}

This article is dedicated to honoring  the memory of Dr. A.F.Andreev who was the true master at finding  new effects and phenomena in hydrodynamics, kinetics  and condensed matter physics, master at formulating  and solving of physical problems by phenomenological methods unrelated to any specific microscopic model.

\bigskip

In materials without space-inversion symmetry there are several specific magneto-electric phenomena and among them the effect of bulk magnetisation
induced by electric current.
The standard relation between the polar vector of electric current density  and the axial vector of density of media magnetisation is ${\bf j}=c~rot~{\bf M}$. However, in 
case of gyrotropic media that is a media without inversion centre  the symmetry allows the direct relationship between the components of  polar and axial vectors
\begin{equation}
M_i=C_{ik}j_k,
\end{equation}
where the matrix $C_{ik}$ depends  of media properties.
So, a flow of current in such a media  induces its magnetisation. This effect was predicted long ago by E.L.Ivchenko and G.E.Pikus \cite{Ivchenko1978} and reviewed in the recently published paper \cite{Ganichev2024}. Particular case of 2D metal with the Rashba spin-orbit interaction was considered in the paper  \cite{Edelstein1990}. 

Recently this phenomenon has been observed in semiconducting tellurium \cite{Furukawa2017} and  in antiferromagnetic metallic compound UNi$_4$B \cite{Amitsuka2018,Izawa2022}. The elemental tellurium forms a trigonal crystal structure P3$_1$21 (D$_3^4)$ or P3$_2$21 (D$_3^6)$ (see Ref.4) without space-inversion. Its point symmetry group is $D_{3d}$. UNi$_4$B has orthorhombic crystal structure Cmcm  with centre of inversion but below transition to an antiferromagnetic state ($T_N\approx 20~K$) apparently transformed to orthorhombic structure Pmm2 with lower symmetry not possessing  space inversion \cite{Sechovsky2021}. Corresponding point symmetry group  is $C_{2v}$.

Magnetic moment found in the paper \cite{Edelstein1990} is directed along vector product of current density on vector $\hat {\bf n}$ of  normal to the 2D plane,  $ C_{ik}\propto e_{ikl}\hat n_l$, and the magnitude of magnetisation
does not depend from the time of scattering on impurities. One gets the impression that we deal with an equilibrium property. In fact, both of these features
follow from an  assumption that the action of electric field comes down to the same shift of momenta of electrons  $\propto e\tau{\bf v}{\bf E}$  in two bands split by the spin-orbit coupling. This is certainly not correct. 
Here, using the kinetic equation for the matrix distribution function of electrons in the band representation  we will derive  a general formula for matrix $C_{ik}$ which is expressed though the intra- and inter-band scattering rates.

\section{Magnetic moment}

 The spectrum of noninteracting
electrons in a metal without inversion center is:
\begin{equation}
\label{H_0}
 \hat \varepsilon
 = \varepsilon\sigma_0+\mbox{\boldmath$\gamma$} 
   \cdot \mbox{\boldmath$\sigma$},
\end{equation}
where 
$\varepsilon=\varepsilon({\bf k})$ denotes the spin-independent part of the spectrum,
$\sigma_0$ is the unit $2\times 2$ matrix in the spin space, $\mbox{\boldmath$\sigma$}=(\sigma_x,\sigma_y,\sigma_z)$ are the Pauli matrices. 
The second term in Eq.
(\ref{H_0}) describes the  spin-orbit  coupling.
The pseudovector $$\mbox{\boldmath$\gamma$}=\mbox{\boldmath$\gamma$}({\bf k})=\gamma_x({\bf k})\hat x+\gamma_y({\bf k})\hat y+\gamma_z({\bf k})\hat z$$  is periodic 
 in the reciprocal space function and         satisfies
$\mbox{\boldmath$\gamma$}(-{\bf k})=-\mbox{\boldmath$\gamma$}({\bf k})$ and 
$g\mbox{\boldmath$\gamma$}(g^{-1} {\bf k})=\mbox{\boldmath$\gamma$}({\bf k})$,
where $g$ is any symmetry operation in the  point group ${\cal G}$ of
the crystal. 

Near the $\Gamma$ point in the case  of $C_{2v}$ point symmetry group the pseudovector $\mbox{\boldmath$\gamma$}({\bf k})$ is  \cite{Samokhin2019}
\begin{equation}
\mbox{\boldmath$\gamma$}({\bf k})=a_1k_y\hat x+a_2k_x\hat y+a_3k_xk_yk_z\hat z.
\end{equation}

The eigenvalues of the matrix (\ref{H_0}) are
\begin{equation}
    \varepsilon_{+}({\bf k})=\varepsilon+\gamma,~~~~~~~~ \varepsilon_{-}({\bf k})=\varepsilon-\gamma,
\label{e3}
\end{equation}
where $\gamma=|\mbox{\boldmath$\gamma$}({\bf k})|$.
The corresponding eigen functions are given by
\begin{eqnarray}
\Psi^+_\sigma({\bf k})=\frac{1}{\sqrt{2\gamma(\gamma+\gamma_z)}}\left (\begin{array} {c}
\gamma+\gamma_z\\
\gamma_+
\end{array}\right),\nonumber\\
~~~~~~~~~~~~\Psi^-_\sigma({\bf k})=\frac{1}{\sqrt{2\gamma(\gamma+\gamma_z)}}
\left(\begin{array} {c}
-\gamma_-\\
\gamma+\gamma_z
\end{array}\right),
\end{eqnarray}
where $\gamma_\pm=\gamma_x\pm i\gamma_y$.
The eigen functions obey the orthogonality conditions
\begin{equation}
\Psi^{\alpha\star}_\sigma({\bf k})\Psi^\beta_\sigma({\bf k})=\delta_{\alpha\beta},~~~~~~~
\Psi^\alpha_{\sigma_1}({\bf k})\Psi^{\alpha\star}_{\sigma_2}({\bf k})=\delta_{\sigma_1\sigma_2}.
\label{ort}
\end{equation}
Here, there is implied the summation over the repeating  spin $\sigma=\uparrow,\downarrow$
or band $\alpha=+,-$ indices.

There are two Fermi surfaces determined by the equations
\begin{equation}
\label{e4}
    \varepsilon_{+}({\bf k})=\mu,~~~~~~~~~~~~\varepsilon_{-}({\bf k})=\mu
\end{equation}
and the Fermi velocities given by the derivatives
\begin{equation}
{\bf v}_{+}=\frac{\partial\varepsilon_{+}({\bf k})}{\partial {\bf k}},~~~~~~~~~~~~~~{\bf v}_{-}=\frac{\partial\varepsilon_{-}({\bf k})}{\partial {\bf k}}~
\end{equation}
taken at points ${\bf k}$ lying at the Fermi surfaces.

The matrix of equilibrium electron distribution function is
\begin{equation}
\hat n
=\frac{n({\varepsilon}_+)+n({\varepsilon}_-)}{2}\sigma_0+\frac{n({\varepsilon}_+)-n({\varepsilon}_-)}{2\gamma} 
\mbox{\boldmath$\gamma$}\cdot \mbox{\boldmath$\sigma$},
\label{eqv}
\end{equation}
where 
\begin{equation}
n({\varepsilon})=\frac{1}{\exp\left(\frac{\varepsilon-\mu}{T}\right)+1}
\end{equation}
is the Fermi function.

The hermitian matrices of the nonequilibrium distribution functions in band and spin representations are related as 
\begin{equation}
f_{\alpha\beta}({\bf k})=\Psi^{\alpha\star}_{\sigma_1}({\bf k})n_{\sigma_1\sigma_2}\Psi^{\beta}_{\sigma_2}({\bf k}).
\label{f}
\end{equation}
In the band representation the equilibrium distribution function  is the diagonal matrix
\begin{equation}
n_{\alpha\beta}=\Psi^{\alpha\star}_{\sigma_1}({\bf k})n_{\sigma_1\sigma_2}\Psi^{\beta}_{\sigma_2}({\bf k})=\left (\begin{array} {cc} n({\varepsilon}_+)&0\\0&n({\varepsilon}_-)  \end{array}\right)_{\alpha\beta}.
\end{equation}
So, the off-diagonal matrix elements $f_{+-}({\bf k},~f_{-+}({\bf k}$ are not equal to zero only out equilibrium.

The kinetic equation for the electron distribution function in non-centrosymmetric metals has been obtained in \cite{Mineev2019} from  the general matrix quasi-classic kinetic equation derived by V.P.Silin \cite{Silin1957}. 
In the presence of external electric field  ${\bf E}$ the corresponding equation for the
stationary deviation of distribution function from equilibrium distribution
\begin{equation}
g_{\alpha\beta}({\bf k})=f_{\alpha\beta}({\bf k})-n_{\alpha\beta}
\end{equation}
is
\begin{eqnarray}
e\left (\begin{array} {cc}({\bf v}_{+}{\bf E}) \frac{\partial n({\varepsilon}_+)}{\partial \varepsilon_+}&({\bf w}_{\pm}{\bf E})(n({\varepsilon}_-)-n({\varepsilon}_+))
\\
({\bf w}_{\mp}{\bf E})(n({\varepsilon}_+)-n({\varepsilon_-}))&({\bf v}_{-}{\bf E})\frac{\partial n({\varepsilon}_-)}{\partial \varepsilon_-}
 \end{array}\right)+
 \left(
\begin{array} {cc}0&i(\varepsilon_--\varepsilon_+)g_{\pm}({\bf k})\\
i(\varepsilon_+-\varepsilon_-)g_{\mp}({\bf k})&0
 \end{array}\right)=\hat I.
 \label{eqv1}
\end{eqnarray}
Here, 
\begin{eqnarray}
{\bf w}_{\pm}({\bf k})=
\Psi^{+\star}_{\sigma}({\bf k})\frac{\partial \Psi^{-}_{\sigma}({\bf k})}{\partial{\bf k}},~~~~~~~~~~~
{\bf w}_{\mp}=-{\bf w}_{\pm}^\star
\label{vel}
\end{eqnarray}
is so called {\bf the Berry connection }and $\hat I$ is the matrix integral of scattering. Unlike to the group velocities ${\bf v}_+$,  ${\bf v}_-$ the dimension of the Berry connections ${\bf w}_\pm$
and ${\bf w}_\mp$ is $1/k$. 

We will limit ourselves to considering the processes of electron scattering on impurities.
In Born approximation the collision integral $I_{\alpha\beta}$ for electron scattering on impurities is (see Appendix A in paper \cite{Mineev2019})
\begin{eqnarray}
I^i_{\alpha\beta}({\bf k})=2\pi n_{imp}\int\frac{d^3k^\prime}{(2\pi)^3}|V({\bf k}-{\bf k}^\prime)|^2\left \{O_{\alpha\nu}({\bf k},{\bf k}^\prime)\left [ g_{ \nu\mu}({\bf k}^\prime)O_{\mu\beta}({\bf k}^\prime,{\bf k})-O_{\nu\mu}({\bf k}^\prime,{\bf k})
  g_{ \mu\beta}({\bf k}) \right ]\delta(\varepsilon^\prime_\nu-\varepsilon_\beta)\right.\nonumber\\ 
 \left .+
  \left[O_{\alpha\nu}({\bf k},{\bf k}^\prime)g_{ \nu\mu}({\bf k}^\prime)-g_{ \alpha\nu}({\bf k})O_{\nu\mu}({\bf k},{\bf k}^\prime)
  \right ]O_{\mu\beta}({\bf k}^\prime,{\bf k})\delta(\varepsilon^\prime_\mu-\varepsilon_\alpha)\right \}.
  \label{matrix1}
\end{eqnarray}
Here, we introduced notations $\varepsilon_{\alpha}=\varepsilon_{\alpha}({\bf k}),~ \varepsilon_{\mu}^\prime=\varepsilon_{\mu}({\bf k}^\prime)$ etc,
\begin{equation}
O_{\alpha\beta}({\bf k},{\bf k}^\prime)=\Psi^{\alpha\star}_\sigma({\bf k})\Psi^\beta_\sigma({\bf k}^\prime)
\end{equation}
and
$
O_{\alpha\beta}({\bf k},{\bf k}^\prime)=O^\star_{\beta\alpha}({\bf k}^\prime,{\bf k}).
$

The solution of Eq.(\ref{eqv1}) has the form:
\begin{eqnarray}
g_{+}=e({\bf v}_{+}{\bf E}) \frac{\partial n({\varepsilon}_+)}{\partial \varepsilon_+}\tau_{+}({\bf k}),\\
g_{-}=e({\bf v}_{-}{\bf E})\frac{\partial n({\varepsilon}_-)}{\partial \varepsilon_-}\tau_{-}({\bf k}),\\
g_{\pm}=e({\bf w}_{\pm}{\bf E})(n({\varepsilon}_-)-n({\varepsilon}_+))\tau_{\pm}({\bf k}),\\
g_{\mp}=e({\bf w}_{\mp}{\bf E})(n({\varepsilon}_+)-n({\varepsilon_-}))\tau_{\mp}({\bf k}),
\end{eqnarray}
where $$\tau_{\pm}({\bf k})=\tau_{\mp}^\star({\bf k}).$$ 
Choosing field direction, say in $x$-direction, and substituting equations (18)-(21) into each matrix element of the matrix kinetic equation (\ref{eqv1}) we come to four field independent equations for four
functions $\tau_{+}({\bf k}),\tau_{-}({\bf k}),\tau_{\pm}({\bf k}),\tau_{\mp}({\bf k})$. 

The density of magnetisation 
\begin{equation}
{\bf M}=\int\frac{d^3{\bf k}}{(2\pi)^3}\mbox{\boldmath$\sigma$}_{\sigma_1\sigma_2}g_{\sigma_2\sigma_1}=\int\frac{d^3{\bf k}}{(2\pi)^3}\mbox{\boldmath$\sigma$}_{\alpha\beta}g_{\beta\alpha}
\end{equation}
is determined by the distribution function and by the Pauli matrices 
in band representation
\begin{equation}
 \mbox{\boldmath$\sigma$}_{\alpha\beta}({\bf k})=\Psi^{\alpha\star}_{\sigma_1}({\bf k}) \mbox{\boldmath$\sigma$}_{\sigma_1\sigma_2}\Psi^{\beta}_{\sigma_2}({\bf k}).
\end{equation}
Here, $\mbox{\boldmath$\sigma$}_{\alpha\beta}({\bf k})=(\sigma^x_{\alpha\beta},\sigma^y_{\alpha\beta},\sigma^z_{\alpha\beta})$ and
\begin{equation}
\sigma^x_{\alpha\beta}=\left (\begin{array} {cc} \frac{\gamma_x}{\gamma}&\frac{-\gamma_-^2+(\gamma+\gamma_z)^2}{2\gamma(\gamma+\gamma_z)}\\
~~&~~\\
\frac{-\gamma_+^2+(\gamma+\gamma_z)^2}{2\gamma(\gamma+\gamma_z)}&-\frac{\gamma_x}{\gamma}\end{array}\right),
~~~~~~
\sigma^y_{\alpha\beta}=\left (\begin{array} {cc} \frac{\gamma_y}{\gamma}&-i\frac{\gamma_-^2+(\gamma+\gamma_z)^2}{2\gamma(\gamma+\gamma_z)}\\
~~&~~\\
i\frac{\gamma_+^2+(\gamma+\gamma_z)^2}{2\gamma(\gamma+\gamma_z)}&-\frac{\gamma_y}{\gamma}\end{array}\right),
~~~~~
\sigma^z_{\alpha\beta}=\left (\begin{array} {cc} \frac{\gamma_z}{\gamma}&-\frac{\gamma_-}{\gamma}\\
~~&~~\\-\frac{\gamma_+}{\gamma}&-\frac{\gamma_x}{\gamma}\end{array}\right).
\end{equation}
Thus, the components of magnetisation are
\begin{equation}
M_x=\int\frac{d^3{\bf k}}{(2\pi)^3}\left [ \frac{\gamma_x}{\gamma}(g_{+}  -g_{-})+\frac{-\gamma_-^2+(\gamma+\gamma_z)^2}{2\gamma(\gamma+\gamma_z)}g_{\mp}
+\frac{-\gamma_+^2+(\gamma+\gamma_z)^2}{2\gamma(\gamma+\gamma_z)}g_{\pm}\right ],
\end{equation}
\begin{equation}
M_y=\int\frac{d^3{\bf k}}{(2\pi)^3}\left [\frac{\gamma_y}{\gamma}(g_{+}  -g_{-})-i\frac{\gamma_-^2+(\gamma+\gamma_z)^2}{2\gamma(\gamma+\gamma_z)}g_{\mp}  +i\frac{\gamma_+^2+(\gamma+\gamma_z)^2}{2\gamma(\gamma+\gamma_z)}g_{\pm}\right ],
\end{equation}
\begin{equation}
M_z=\int\frac{d^3{\bf k}}{(2\pi)^3}\left [\frac{\gamma_z}{\gamma}(g_{+}  -g_{-}) -\frac{\gamma_-}{\gamma}g_{\mp} -\frac{\gamma_+}{\gamma}g_{\pm}  \right ].
\end{equation}
Substituting equations (18)-(21) into expressions (25)-(27) we see that electric field of arbitrary direction induces the magnetisation along all three crystallographic axis:
\begin{equation}
M_i=A_{ij}E_j.
\label{E}
\end{equation}
In the absence of the band splitting caused by spin-orbit coupling $g_+=g_-$ and $g_\pm=g_\mp=0$, we come to ${\bf M}=0$ as it should be in a material with centre of inversion.

To find a relationship of magnetisation and electric current one needs to determine the corresponding filed-current relation.
  The electric  current density is determined by the following expression
 \begin{equation}
 {\bf j}=e\int \frac{d^3k}{(2\pi)^3}\frac{\partial \varepsilon_{\sigma\sigma_1}({\bf k})}{\partial {\bf k}}g_{\sigma_1\sigma}({\bf k},\omega).
 \end{equation}
Transforming it to the band representation we get
\begin{widetext}
\begin{eqnarray}
 {\bf j}=e\int \frac{d^3k}{(2\pi)^3} \Psi_\sigma^{\alpha\star}({\bf k})\frac{\partial \varepsilon_{\sigma\sigma_1}({\bf k})}{\partial {\bf k}}
 \Psi_{\sigma_1}^{\gamma}({\bf k})\Psi_{\sigma_2}^{\gamma\star}({\bf k})
 g_{\sigma_2\sigma_3}({\bf k},\omega)\Psi_{\sigma_3}^\alpha({\bf k})\nonumber\\
 =e\int \frac{d^3k}{(2\pi)^3} \left \{
 \frac{\partial \varepsilon_{\alpha\gamma}({\bf k})}{\partial {\bf k}}
 +\left [ \Psi_{\sigma}^{\alpha\star}({\bf k})
 \frac{\partial\Psi^{\beta}_{\sigma}}{\partial {\bf k}},  \varepsilon_{\beta\gamma} \right]\right\}
 g_{\gamma\alpha}({\bf k},\omega),
\end{eqnarray}
\end{widetext}
where $\left [\dots,\dots\right ]$ is the commutator. Performing matrix multiplication we obtain
\begin{equation}
{\bf j}=e\int \frac{d^3k}{(2\pi)^3} \left [ {\bf v}_+g_++{\bf v}_-g_-+({\bf w}_{\pm}g_{\mp}- {\bf w}_{\mp}g_{\pm})(\varepsilon_--\varepsilon_+)  \right ].
\end{equation}
Let us note that unlike to the magnetisation the electric current does not equal to zero in the absence of the band splitting.

Substituting here  the equations (18)-(21) we come to the linear relation between the components of electric current and  applied field
\begin{equation}
j_k=\sigma_{kl}E_l,
\label{J}
\end{equation}
or
\begin{equation}
E_j=\rho_{jk}j_k,~~~~~~~~~~~\hat\rho=\hat\sigma^{-1},
\end{equation}
which together with Eq.(\ref{E}) presents the relation we are searching for
\begin{equation}
M_i=A_{ij}\rho_{jk}j_k.
\end{equation}
The matrices $A_{ij}$ and $\rho_{jk}$ are expressed through four  momentum dependent time of scattering $\tau_+({\bf k}),\tau_{-}({\bf k}),\tau_{\pm}({\bf k}),\tau_{\mp}({\bf k})$ determined by four integral equations explicitly expressed through the components of pseudo-vector of spin-orbit interaction.
If we formally put all the scattering times as momentum independent and equal to each other $\tau_+=\tau_{-}=\tau_{\pm}=\tau_{\mp}=\tau$  we come to the expression for
magnetisation independent of time of scattering  $\tau$ as it was found in the paper \cite{Edelstein1990}.

\section{Conclusion}
By the direct calculation based on the application of kinetic equation for the matrix distribution function of electrons occupying the states in two bands split by the spin-orbit interaction there was demonstrated that electric current flowing through a nonmagnetic metal with crystal structure without inversion center induces magnetisation. Component of induced magnetic moment are expressed as functionals from momentum dependent components of pseudovector of spin-orbit coupling particular for each crystallographic symmetry.

In UNi$_4$B below phase transition to an antiferromagnetic state $(T_N\approx 20 K)$ \cite{Amitsuka2018,Izawa2022} accompanying by the breaking of the space parity \cite{Sechovsky2021}   is appeared the magnetisation proportional to electric current. The magnitude of induced magnetisation increases with temperature decrease what corresponds to growing of amplitude of antiferromagnetic ordering. The established experimentally temperature dependence of magnetisation \cite{Izawa2022}  looks as $M(T)\propto (T_N-T )^\beta$ with $\beta > 1$. This type behaviour is qualitatively understandable but cannot be derived in frame developed theory because at $T\to T_N$  the band splitting due to breaking of space parity becomes less than the scattering rate of electrons on lattice defects, impurities and phonons $\gamma(T)|_{T\to T_N}<1/\tau$
and the model with so small band splitting is applicable only qualitatively.

\bigskip
{\bf Acknowledgment}
I am grateful to Dr.Koichi Izawa, whose report presented on the French-Japanese Symposium in October 2023 stimulated my interest to this subject.
I am also grateful to Dr. E.L. Ivchenko, who introduced me to previously unknown to me publications on this subject.


\begin{thebibliography}{220}


\bibitem{Ivchenko1978} E.L.Ivchenko, G.E.Pikus, Pis'ma  ZhETF {\bf 27}, 640 (1978) [JETP Letters {\bf 27}, 604 (1978)].

\bibitem{Ganichev2024}S.D.Ganichev, E.L.Ivchenko, Encyclopedia of Cond. Mat. Phys. (Second Ed.) {\bf 2}, 177 (2024).

\bibitem{Edelstein1990}V.M.Edelstein, Sol. St. Comm. {\bf 73}, 233 (1990).

\bibitem{Furukawa2017}Tetsuya Furukawa, Yuri Shimokawa, Kaya Kobayashi,Tetsuaki Itou, Nat.Commun. {\bf 8}, 954 (2017).

\bibitem{Amitsuka2018} Hiraku Saito, Kenta Uenishi, Naoyuki Miura, Chihiro Tabata, Hiroyuki Hidaka, Tatsuya Yanagisawa, and Hiroshi Amitsuka,	
J. Phys. Soc. Jpn. {\bf 87}, 033702 (2018).

\bibitem{Izawa2022}K. Ota, M. Shimozawa, T. Muroya, T. Miyamoto, S. Hosoi,
A. Nakamura, Y. Homma, F. Honda, D. Aoki, and K. Izawa, arXiv:2205.05555v1[cond-mat.str-el].

\bibitem{Sechovsky2021}J. Willwater, S. S{\"u}low, M. Reehuis, R. Feyerherm, H. Amitsuka, B. Ouladdiaf, E. Suard, M. Klicpera, M. Vali{\v s}ka, J. Pospı{\v s}il and V. Sechovsky, Phys. Rev. B {\bf 103}, 184426 (2021).

\bibitem{Samokhin2019}K.V.Samokhin, Annals Phys.{\bf 407}, 179 (2019).

\bibitem{Mineev2019}V.P.Mineev, Zh. Exp. Teor. Fiz. {\bf 156}, 750 (2019); Erratum {\bf 157}, 1131 (2020) [JETP {\bf 129}, 700 (2019); Err. {\bf 130}, 955 (2020)] .

\bibitem{Silin1957} V.P.Silin, Zh.  Eksp.Teor.Fiz.  {\bf 33}, 1227 (1957) [Sov.  Phys. JETP {\bf 6}, 945 (1958)]. 

\end{thebibliography}
\end{document}